\documentclass[nospthms,reqno]{svjour2}

\smartqed

\usepackage{amssymb}
\usepackage{mathrsfs}
\usepackage{dsfont}
\usepackage{amsmath,amsthm}
\usepackage{amsfonts}
\usepackage{enumerate, xspace}
\usepackage{color}
\RequirePackage{hyperref}
\usepackage{changebar}
\usepackage{hyperref}
\usepackage{pdfsync}
\usepackage{graphicx}
\usepackage{a4wide}

\numberwithin{equation}{section}

\newtheorem{thm}{Theorem}[section]

\newtheorem{rem}[thm]{Remark}

\def\si{{\sigma}}

\def\la{{\lambda}}
\def\La{{\Lambda}}

\def\om{{\omega}}

\title{A note on Fick's law with phase transitions}

\titlerunning{Fick's law with phase transitions}

\author{
Anna De Masi \and
Stefano Olla \and Errico Presutti }

\institute{ Anna De Masi \at
               Dipartimento di Ingegneria e Scienze dell'Informazione e Matematica, Universit\`a di L'Aquila
 L'Aquila, 67100 Italy 
              \email{demasi@univaq.it}  
              \and {Stefano Olla \at
CEREMADE, UMR-CNRS, Universit\'e de Paris Dauphine, PSL Research University, 75016 Paris, France
\email {olla@ceremade.dauphine.fr}
\and Errico Presutti \at GSSI, viale F. Crispi 7, 67100 L'Aquila, Italy
 \email{errico.presutti@gmail.com}   }}

\begin{document}

\maketitle

\begin{abstract} We characterize the non equilibrium stationary states in two classes of systems where phase transitions are present. We prove that the interface in the limit is a plane which separates the two phases.
\end {abstract}

\section{Introduction}
\label{sec.1}

While the hydrodynamic limit of ``closed'' diffusive particle systems is well  established, less is known for ``open systems'' where boundary processes are added to the bulk evolution. The aim is to characterise the invariant measures which carry a constant particles flux through the system (non equilibrium stationary states) and prove the validity of the Fick law which states that the particles current scales as the inverse of the size of the system.  Our main purpose is to check the validity of the above picture when phase transitions are present.

Hydrodynamic limit in a torus when phase transitions are present has been first established by Rezakhanlou \cite{R} for a stochastic lattice Ginzburg-Landau model, see Section \ref{sec.2} below, and then by Bertini, Butt\`a and R\"udiger, \cite{BBR}, for a system where Ising spins and Ginzburg-Landau variables are coupled, see  Section \ref{sec.3} below.  The  analysis in  these papers is based on entropy inequalities as  introduced by Guo, Papanicolaou and Varadhan \cite{GPV}.
An essential point in these two papers is that the current of the conserved quantity $m$ is
  the gradient of the chemical potential: this is important because the chemical potential is continuous at the interface between the different phases.
The above systems have unbounded configuration space, and the tightness in the proof of the hydrodynamic limit requires bounds on the moments of the currents that are usually proven via entropy inequality.
  This creates a difficulty in extending the analysis  to open systems with boundaries which are in contact with reservoirs and a proof of convergence in the hydrodynamic limit is still an open problem.

The limit diffusive equation in closed systems has the form
   $$
   \frac{\partial m}{\partial t} = \Delta \la
   $$
where $\la$ is the chemical potential. Thus linear chemical potential profiles
are stationary and if the above equation retains its validity also in open systems we would get a characterization of the stationary profiles simply as
profiles where the chemical potential varies linearly in space. In this paper we  check that in the two systems quoted above, see Section \ref{sec.2} and \ref{sec.3},  it is indeed true that the chemical potential varies linearly: this happens not only in the hydrodynamic limit but also when the system is finite. In fact we compute explicitly the stationary measure proving that it is what is called a local equilibrium state with a slowly varying chemical potential.
In particular we see that the stationary density profiles at phase transition are discontinuous: a planar interface separates regions where the density is  ``in the plus'' and in the ``minus'' phases. The limit density profile is then a solution of the stationary free boundary Stefan problem.
We have not found in the literature such statements and in this short note we report a proof.

 The stationary non equilibrium state in our models has short range correlations
  contrary to what is generally believed, \cite{DLS,DLS2}.
 According to fluctuating hydrodynamic theory \cite{derrida}, and non-equilibrium thermodynamic \cite{mft}, short-range correlations in stationary non equilibrium states occur when  the mobility and the diffusion coefficient are suitably related. Such relation is satisfied in the zero range processes, \cite{DF}, as well as in  our models where  the mobility is equal to one.

\section{The stochastic Ginzburg-Landau model}
\label{sec.2}
We consider a Ginzburg-Landau stochastic lattice model in the region $\La= [1,N]^d$ of $\mathbb Z^d$, $N>1$. We denote by $x$ the points in $\La$ and by $e_i$ the unit vector in the $i$-th direction, $i=1,..,d$.  $\phi_x \in \mathbb R$ is ``the charge'' at site $x\in \La$ and $\phi=\{\phi_x, x\in \La\}$, a charge configuration. Since $\La$ will be kept fixed we do not make explicit the dependence on $\La$. The Hamiltonian we consider is
\begin{equation}
\label{2.1}
H(\phi)= \sum_x (\phi_x ^2 - 1)^2 +  \sum_{x}\sum_{i=1,..,d}\mathbf 1_{x+e_i\in \La}(\phi_x   - \phi_{x+e_i})^2.
\end{equation}
For $\beta$ large enough and $d \ge 2$ the corresponding Gibbs measure at infinite volume has a phase transition with the two extremal translation invariant Gibbs measures which are supported respectively by configurations where $\phi_x$ is
predominantly close either to $+1$ or to $-1$, \cite{Dyn-Sinai}.

What follows holds as well in more general cases as: vertical periodic conditions
($x+Ne_i$ being identified to $x$, $i\ge 2$) and when the Hamiltonian includes interactions with an external configuration; we can also allow for more general one body potentials and finite range interactions, provided the Hamiltonian remains super-stable (\cite{LP}, definition 1.4).
For notational simplicity we restrict to the Hamiltonian $H(\phi)$ of \eqref{2.1}.

We denote by $\nu_\la(d\phi)$ the Gibbs measure with ``chemical potential'' $\la \in \mathbb R$:
\begin{equation}
\label{2.2}
\nu_\la(d\phi) = Z_\la^{-1} e^{-\beta [H(\phi)-\la \sum_x \phi_x]} d\phi, \quad d\phi=\prod_{x\in \La} d\phi_x
\end{equation}
and write $\nu$ for $\nu_0$ (i.e.\ when the chemical potential is $0$).

Dynamics is defined by the generator
\begin{equation}
\label{2.3}
L =  \sum_{x}\sum_{i=1,..,d}\mathbf 1_{x+e_i\in \La} L_{x,x+e_i}
+ \sum_{x: x\cdot e_1=1} B_x + \sum_{x: x\cdot e_1=N} B_x
\end{equation}
where $ L_{x,x+e_i}$ and $B_x$ are defined as follows.
\begin{equation}
\label{2.4}
L_{x,x+e_i} = -\Big(\frac{\partial H}{\partial \phi_x}
- \frac{\partial H}{\partial \phi_{x+e_i}}\Big)\Big(\frac{\partial }{\partial \phi_x}
- \frac{\partial }{\partial \phi_{x+e_i}}\Big) + \frac 1 \beta
\Big(\frac{\partial }{\partial \phi_x}
- \frac{\partial }{\partial \phi_{x+e_i}}\Big)^2
\end{equation}
and given $\la_-$ and $\la_+$ in $\mathbb R$:
\begin{equation}
\label{2.5}
B_x = -\Big(\frac{\partial H}{\partial \phi_x}
- \la_+\Big) \frac{\partial }{\partial \phi_x}
  + \frac 1 \beta
\Big(\frac{\partial }{\partial \phi_x}
\Big)^2,\quad x:x\cdot e_1=N
\end{equation}
\begin{equation}
\label{2.6}
B_x = -\Big(\frac{\partial H}{\partial \phi_x}
- \la_-\Big) \frac{\partial }{\partial \phi_x}
  + \frac 1 \beta
\Big(\frac{\partial }{\partial \phi_x}
\Big)^2,\quad x:x\cdot e_1=1 .
\end{equation}
Observe that $L_{x,x+e_i}$ is self-adjoint with respect to $\nu_\la$ for any $\la \in \mathbb R$, $B_x$, $x\cdot e_1=N$, is self-adjoint with respect to $\nu_{\la_+}$ and $B_x$, $x\cdot e_1=1$,   with respect to $\nu_{\la_-}$.

We denote by $L^*$, $L^*_{x,x+e_i}$ and by $B^*_x$ the adjoints of $L$, $L_{x,x+e_i}$ and   $B_x$  with respect to $\nu$, then, as already mentioned, $L^*_{x,x+e_i} = L_{x,x+e_i}$ while
\begin{equation}
\label{2.7}
B^*_x = e^{\beta \la_{\pm} \sum_y \phi_y} B_x e^{-\beta \la_{\pm} \sum_y \phi_y},
\quad x\cdot e_1=N,\;\;\text{respectively}\;\;x\cdot e_1=1 .
\end{equation}

Let $\mu$ be a probability absolutely continuous with respect to $\nu$
and write $\mu= f \nu$. If $L^*f=0$  then $\mu$ is time invariant.

\begin{thm}
\label{thm2.1}
The measure $\mu=f\nu$ is invariant if
\begin{equation}
\label{2.8}
 f= e^{\beta\sum_x \la_x \phi_x}, \quad \la_x = \la(x\cdot e_1)
\end{equation}
where
\begin{equation}
\label{2.9}
\la(i)= \la_- + \frac{ \la_+-\la_-}{N+1}\, i
\end{equation}

\end{thm}

\medskip
\noindent
{\bf Proof.} We fix hereafter $\la_x$ according to  \eqref{2.8}--\eqref{2.9}.  We have $L^*_{x,x+e_i} f=0$ if $i>1$, and for $x,x+e_1 \in \La$, calling $i=x\cdot e_1$,
\begin{equation}
\label{2.10}
L^*_{x,x+e_1}f= f\Big\{ \big(\frac{\partial H}{\partial \phi_{x+e_1}}-\frac{\partial H}{\partial \phi_x}
 \big)\big( \la(i)-\la(i+1)\big)
+ \frac 1 \beta \big(\la(i+1)-\la(i)\big)^2\Big\}
\end{equation}
Using \eqref{2.7}, for $x:x\cdot e_1=N$:
\begin{equation}
\label{2.11}
B^*_{x}f= f\Big\{ - \big(\frac{\partial H}{\partial \phi_{x}}-\la_{+}\big)(\la(N)-\la_+)
+ \frac 1 \beta \big(\la(N)-\la_+\big)^2\Big\}
\end{equation}
Similarly, for $x:x\cdot e_1=1$:
\begin{equation}
\label{2.12}
B^*_{x}f= f\Big\{ - \big(\frac{\partial H}{\partial \phi_{x}}-\la_{-}\big)(\la(1)-\la_-)
+ \frac 1 \beta \big(\la(1)-\la_-\big)^2\Big\}
\end{equation}
After a telescopic cancellation it then follows that $L^*f=0$.  \qed

\begin{rem}
  Uniqueness of this stationary measure follows from the hypoellipticity of $L$, see e.g. \cite{reybellet}.

\end{rem}

\section{The stochastic phase-field model}
\label{sec.3}

We use the same notation as in the previous model adding as new variables
the Ising spins $\si_x \in \{-1,1\}$,  $\si= \{\si_x, x \in \La\}$, being the
spin configuration in $\La$.  The Hamiltonian  $H(\si,\phi)$ is
\begin{equation}
\label{3.1}
H(\si,\phi) =   - \sum_{x } \sum_{i=1,..,d}\mathbf 1_{x+e_i\in \La}\si_x\si_{x+e_i}+\frac 12 \sum_x \phi_x^2
\end{equation}

The generator  is
\begin{equation}
\label{3.2}
L = \sum_{x } L_{x} + \sum_{x}\sum_{i=1,..,d}\mathbf 1_{x+e_i\in \La} L_{x,x+e_i}
+ \sum_{x: x\cdot e_1=1} B_x + \sum_{x: x\cdot e_1=N} B_x
\end{equation}
where
\begin{equation}
\label{3.3}
L_x g(\si,\phi) = c_x(\si,\phi) [ g((\si,\phi)^x)-g(\si,\phi)],\quad
 c_x(\si,\phi) = e^{-\frac \beta 2 [ H((\si,\phi)^x)-H(\si,\phi)]}
\end{equation}
with $(\si,\phi)^x(x) = (-\si(x), \phi(x)+2\si(x))$ and $(\si,\phi)^x(y)=(\si,\phi)(y)$ for $y\ne x$.  As in the previous section (but with the new Hamiltonian)
\begin{equation}
\label{3.4}
L_{x,x+e_i} = -(\phi_x-\phi_{x+e_i})(\frac{\partial}{\partial \phi_x}
-\frac{\partial} {\partial \phi_{x+e_i}} )+ \frac 1 \beta
(\frac{\partial}{\partial \phi_x}
-\frac{\partial }{\partial \phi_{x+e_i}} )^2
\end{equation}
As in the previous section (but with the new Hamiltonian)
\begin{equation}
\label{3.5}
B_x = -\Big( \phi_x
- \la_+\Big) \frac{\partial }{\partial \phi_x}
  + \frac 1 \beta
\Big(\frac{\partial }{\partial \phi_x}
\Big)^2,\quad x:x\cdot e_1=N
\end{equation}
and
\begin{equation}
\label{3.6}
B_x = -\Big(\phi_x
- \la_-\Big) \frac{\partial }{\partial \phi_x}
  + \frac 1 \beta
\Big(\frac{\partial }{\partial \phi_x}
\Big)^2,\quad x:x\cdot e_1=1
\end{equation}


\noindent
{\bf Remarks.}

\begin{itemize}

\item
The quantity $\om (x) := \si(x) + \phi(x)$ is invariant under the  bulk evolutions (i.e.\ without the generators $B_x$).

\item  For $\beta$ large the Ising part of the Hamiltonian   has a phase transition when $d\ge 2$.

\item  Let $f=f(\si)$ (i.e.\ it does not depend on $\phi$) then
\begin{equation}
\label{2.11a}
L f(\si) = \sum_x e^{-\frac \beta 2 [ H^{\rm ising}_N(\si^x)-H^{\rm ising}_N(\si)
+2\si_x\phi_x]} [ f(\si^x)-f(\si)]
\end{equation}
($H^{\rm ising}_N$ the Ising Hamiltonian). Thus the spin $\si_x$ is updated using the Glauber dynamics with additional magnetic field $\phi_x$.  On the other hand (under the bulk dynamics)
\begin{equation}
\label{2.12a}
\frac{d}{dt} E_{\sigma_0, \phi_0}[\phi_t(x)] =  E_{\sigma_0, \phi_0}[ \Delta\phi_t(x)] -
\frac{d}{dt} E_{\sigma_0, \phi_0}[\si_t(x)]
\end{equation}
where $\Delta$ is the discrete Laplacian and $E_{\sigma_0, \phi_0}$
denotes the expectation with respect to the dynamics generated by $L$
with initial contitions $({\sigma_0, \phi_0})$.
This is therefore the discrete stochastic version of the phase field dynamics in continuum mechanics, see \cite{BBR} for comments.

\end{itemize}

\noindent
We denote by $d\nu_\la(\si,\phi)$ the
Gibbs measure
with ``chemical potential'' $\la \in \mathbb R$:
\begin{equation}
\label{3.7}
d\nu_\la(\si,\phi) = Z_\la^{-1} e^{-\beta [H(\si,\phi)-\la \sum_x (\si_x+\phi_x)]} d\phi
\end{equation}
and write again $\nu$ for $\nu_0$.

As in \autoref{sec.2},  $L_{x,x+e_i}$ is self-adjoint with respect to $\nu_\la$ for any $\la \in \mathbb R$, while $B_x$, $x\cdot e_1=N$, is self-adjoint with respect to $\nu_{\la_+}$ and $B_x$, $x\cdot e_1=1$,   with respect to $\nu_{\la_-}$.  The generators $L_x$ are self-adjoint with respect to
\begin{equation}
\label{3.8}
e^{\sum_y \la_y (\si_y+\phi_y)}d\nu(\si,\phi)
\end{equation}
for any choice of $\la_y$ and therefore
\begin{equation}
\label{3.9}
 L^*_x e^{\sum_y \la_y (\si_y+\phi_y)}=0
\end{equation}
We choose $\la_x$ as in \eqref{2.8}--\eqref{2.9} and then using \eqref{2.10}--\eqref{2.11}--\eqref{2.12} we see that Theorem \ref{thm2.1} extends its validity to the present case, details are omitted.

\section{Hydrodynamic limit}
\label{sec.4}

In this section we study the limit as $N\to \infty$ (hydrodynamic limit)
of the stationary measures found in the previous sections. It is now convenient to underline the dependence on $N$ and we thus write $\La^{(N)}$ for $\La$ and $\mu^{(N)}$ for the corresponding stationary measures.  Denote by $C= [-\frac 12,\frac 12]^d$ the unit cube of $\mathbb R^d$ centered at the origin, so that $C$ is obtained in the limit $N\to \infty$ by squeezing $\La^{(N)}$ by $N$ and then shifting by $-\frac 12 e_1$.  We call $\{g\}$ the set of $C^\infty$ test functions on $C$.

In the phase field Ising system of the previous section we have a full description of the hydrodynamic limit, for the Ginzburg-Landau model we need some assumptions on the behavior of the invariant measures $\mu^{(N)}$.  We start from the former and introduce the magnetization fields $X(g)$:
\begin{equation}
\label{4.1}
 X(g)= N^{-d} \sum_{x\in \La^{(N)}} g\left( \frac xN - \frac 12 e_1\right) \si_x
\end{equation}
We then have:

\begin{thm}
\label{thm4.1}
For any test function $g$ and any $\delta>0$
\begin{equation}
\label{4.2}
\lim_{N\to \infty}\mu^{(N)} \Big[ | X(g)- \int_C dr\,
g(r)m(r)|> \delta\Big] =0
\end{equation}

where, denoting by $\nu_h$, $h \ne 0$, the Ising DLR measure with external magnetic field $h$:
\begin{equation}
\label{4.3}
m(r) = E_{\nu_{h(r)}}[\si_0], \quad h(r)=  \la_- + (\la_+-\la_-) [r\cdot e_1 + \frac 12],\qquad r: h(r)\ne 0
\end{equation}

\end{thm}

\medskip
\noindent
{\bf Proof.}  Let $\Delta\subset \mathbb Z^d$ be a cube of side $\ell$ with center $x^*(N)$
such that
    $$
    \lim_{N\to \infty} \frac{x^*(N)}{N} = r+\frac 12  e_1,\quad (r-r_0) \cdot e_1 \ne 0, r \notin \partial C
    $$
where $r_0$ is such that $h(r_0)=0$, if there is no such point the condition is dropped (it can also be dropped if $\beta<\beta_c$, the inverse critical temperature).

 We are going to prove that,  for any $\delta>0$,
\begin{equation}
\label{4.4}
 \lim_{\ell\to \infty}  \limsup_{N\to \infty} \;
\mu^{(N)}\Big[\Big| \frac{1}{|\Delta|} \sum_{x\in \Delta} \si(x) - m(r)\Big| >\delta   \Big] =0,
\end{equation}
 then \eqref{4.2}  follows from \eqref{4.4}.

To prove \eqref{4.4} we first observe that for any $x$ in $\Delta$
    $$
    |\la_x - h(r)| \le c \frac{\ell}{N}
    $$
We then
condition on the spins outside $\Delta$ and call $\mu^{(N)}_{\si_{\Delta^c}}(\si_\Delta)$
the probability of having $\si_\Delta$ when there is  $\si_{\Delta^c}$ outside $\Delta$.  Analogously we call
$\nu_{\Delta,h(r); \si_{\Delta^c}}(\si_\Delta)$ the Ising Gibbs  probability of having $\si_\Delta$ in $\Delta$ when there is an external magnetic field $h(r)$ and when the boundary conditions are $\si_{\Delta^c}$.
We have
\begin{equation}
\label{4.5}
 e^{-c' \ell^d\frac{\ell}{N}} \le\frac{\mu^{(N)}_{\si_{\Delta^c}}(\si_\Delta)}{\nu_{\Delta,h(r); \si_{\Delta^c}}(\si_\Delta)} \le e^{c' \ell^d\frac{\ell}{N}}
\end{equation}
We also have for all $\si_\Delta$ and $ \si_{\Delta^c}$
\begin{equation}
\label{4.6}
\nu_{\Delta,h(r); \si_{\Delta^c}}(\si_\Delta) \le e^{c'' \ell^{d-1}}
\nu_{h(r)}(\si_\Delta)
\end{equation}
where $\nu_{h(r)}(\si_\Delta)$ denotes the probability of having the configuration
$\si_\Delta$ in $\Delta$ with respect to the infinite volume DLR measure $\nu_{h(r)}$ \cite{P}.
We then use the large deviation estimate (valid when $\beta<\beta_c$ and otherwise when the external magnetic field is non-zero)
\begin{equation}
\label{4.7}
 \nu_{h(r)}\Big[\Big| \frac{1}{|\Delta|} \sum_{x\in \Delta} \si(x) - m(r)\Big| >\delta   \Big] \le e^{-c''' \delta \ell^d}
\end{equation}
and conclude the proof of \eqref{4.4}.  \qed

Suppose $\beta<\beta_c$, then the relation \eqref{4.3} between the magnetization $m$ and the magnetic field $h$ can be inverted. Let $F(m)$ be the thermodynamic free energy density at the value $m$ of the magnetization density.  Then $m  = E_{\nu_{h}}[\si_0]$ ranges in $(-1,1)$ when $h\in \mathbb R$ and
\begin{equation}
\label{4.8}
m  = E_{\nu_{h}}[\si_0], \quad h =F'(m)\equiv \frac{dF(m)}{dm}
\end{equation}
Then Theorem  \ref{thm4.1} states that $ F'(m(r))$ is a linear function of $r$ so that
\begin{equation}
\label{4.9}
\frac{d^2}{dr^2} F'(m(r)) =0
\end{equation}
\eqref{4.9} agrees with the analysis in \cite{BBR} where it is shown that the hydrodynamic limit (in the torus) is a diffusive equation with diffusion coefficient $D(m)= F''(m)$.

The most interesting case  is when there is a phase transition, we thus restrict below to $d \ge 2$ and  $\beta>\beta_c$.  The relation \eqref{4.3}  between $m$ and $h$ is then no longer invertible.  We still have a one to one correspondence between $m$ and $h$ when $h \in \mathbb R\setminus 0$ and $m: |m| > m_\beta$, $m_\beta>0$ the spontaneous magnetization but there is no value of $h$ for which $|m| < m_\beta$.  For any $h\ne 0$ there is a unique DLR measure and \eqref{4.3} holds with $m$ such that $|m| > m_\beta$.  $m$ converges to $\pm m_\beta$ as $h \to 0^+$ and respectively to $0^-$, if $h=0$ there are two pure states, one with magnetization $m_\beta$ and the other with magnetization $-m_\beta$.

Thus if there is $r_0$ so that $h(r_0)=0$, then the magnetization profile $m(r)$ of Theorem \ref{thm4.1} has a discontinuity on the vertical plane $r: r\cdot e_1=r_0\cdot e_1$ with a jump from $-m_\beta$ to $+m_\beta$ when moving from negative to positive values of the magnetic field.

For the sake of definiteness let us suppose hereafter that $\la_+>0$ and $\la_- = -\la_+$ so that the [macroscopic] interface is localized in the plane $r\cdot e_1 =0$.  The question is about the microscopic location of the interface.  This problem has been much studied in equilibrium when the interface is determined by the boundary conditions.  The typical case is when we put $+$ boundary conditions on $\La^c \cap \{x\cdot e_1 \ge 0\}$ and  $-$ boundary conditions on $\La^c \cap \{x\cdot e_1 < 0\}$, i.e.\ on the left  and right semi-spaces.  In this case the interface fluctuates around the plane $x\cdot e_1=0$  by the order of $\sqrt N$ in $d=2$ while, if $\beta$ is large enough when $d\ge 3$ it becomes rigid, giving rise, in the thermodynamic limit to a non translation invariant DLR measure known as the ``Dobrushin state'' \cite{D}.
We believe that also in our case the interface is rigid in $d\ge 3$ but a proof would go beyond the purposes of this short note.  In $d=2$ the question remains open whether  the interface fluctuates as $\sqrt N$ as in the equilibrium setting or, as we believe, its fluctuations are damped by the presence of the chemical potential. Fluctuations have also being considered in \cite{Spohn}.

Another interesting phenomenon appears when we let $\la_+ \to 0$ as $N\to \infty$.  If  $\la_+ \to 0$ slowly enough the limit profile $m(r)$ is a step function equal to $-m_\beta$ and to $m_\beta$ for $r\cdot e_1<0$ and respectively  $r\cdot e_1>0$.  However we see a different picture if the convergence is fast enough and supposing that the Ising Hamiltonian $H$ has no interaction with the outside (zero boundary conditions).  The limit profile is in fact random and constantly equal to  $m_\beta$ with probability $1/2$ and to $-m_\beta$ with same probability.  Indeed the energy cost of an interface scales as $N^{d-1}$, the state with the plus phase in the whole region, i.e.\ without any interface, has also a cost, namely the energy of having the plus phase ($m_\beta$) also in $r\cdot e_1 <0$.  This is bounded proportionally to $|\la_-| N^d$ thus if $\la_- = c N^{-a}$, $a>1$, $|\la_-|N \to 0$ and therefore the probability of an interface vanishes.

\subsection {The Ginzburg-Landau case}  The conjecture is that the analogue of Theorem \ref{4.1} holds, there are however difficulties when extending the proof given for Ising. In particular \eqref{4.5} and \eqref{4.6} cannot hold uniformly on  the boundary conditions. 
As done by Rezakhanlou, \cite{R},  the problem can be circumvented by cutoffing the interaction term
$(S_x - S_{x+e})^2$.  However we could not quote anymore the literature for a proof of phase transition and the validity of the Pirogov-Sinai scheme (see \cite{P}) to prove absence of phase transitions for $\la \ne 0$.

\section{Boundary conditions}
\label{sec.5}

The Fick's law is about the current which flows in a system having fixed
the density at the boundaries.
In the hydrodynamic limit literature this is usually enforced by
fixing the chemical potential (as we did in Sections \ref{sec.2} and \ref{sec.3}), technically this is used to prove Gibbsian local equilibrium up to the boundaries.  The two procedures, fixing the density or fixing the chemical potential, are equivalent in the absence of phase transitions.  At phase transition instead there is an interval of values of the density (the spinodal region) which are not produced by any chemical potential, the question then arises on how such densities can be enforced and which are the consequences.

Fixing the chemical potential has a clear physical meaning.  Consider in fact two systems separated by a ``wall'' which however allows for exchange of particles.  Equilibrium is then reached when the chemical potentials at the wall are equal.
The free energy flow from left to right at the wall is
$$
\Big(\frac{dF_{\rm left}(m)}{dm} -\frac{dF_{\rm right}(m)}{dm}  \Big)\;j
$$
$j$ the current at the wall.  Since the derivative of the free energy is the chemical potential the above is equal to 0.

Thus fixing the chemical potential means physically that there is no free energy dissipation at the boundaries.  As a consequence if we force a boundary process which fixes the density at the boundary in the spinodal region we should expect some extra free energy flux at the boundary.  Since the whole process (in the Fick's law apparatus) is ``boundary driven'' such extra flux may not be innocent at all.  Indeed numerical simulations performed on the $d=2$ Ising model with Kawasaki dynamics show a very rich behavior including the fact that the current may flow from the boundary with smaller density to the one with larger density, phenomenon known as ``uphill diffusion'', \cite{CDP,CDP2,CGGV}.

\section*{Acknowledgments}

We are indebted to B. Derrida, D. Gabrielli and J. L. Lebowitz for many useful comments on the nature of the stationary non equilibrium states.

S.O. would like to thank the warm hospitality of the GSSI, where this work was initiated.
The work of S.O. has been partially supported by the grant ANR-15-CE40-0020-01 LSD 
of the French National Research Agency.

\end{document}